%% file: luciv_STM_2014_b5.tex
\documentclass[b5paper,twoside]{stm14b5}

\usepackage[pdftex]{graphicx} 

\usepackage[utf8]{inputenc}
\usepackage[T1]{fontenc}
\usepackage{euscript}
\usepackage{amsfonts,amssymb,amsmath}
\usepackage{amsthm}
\usepackage{feynmp}
\usepackage{enumerate}
\usepackage{amscd}

\usepackage{hyperref}

\def\url#1{\texttt{#1}}

\usepackage{bm}
\usepackage{epsf}

\def\mv{{\bm v}}

\def\S{\mathcal{S}}

\def\dRM{{\mathrm d}}
\def\mk{{\bm k}}

\def\mx{{\bm x}}

\def\boldnabla{{\bm \nabla}}

\newcommand{\NB}[1]{\ensuremath{\biggl( #1  \biggl) }}

\begin{document}

\setcounter{equation}{0}
\setcounter{figure}{0}
\setcounter{section}{0}

\input{lucivjansky_14.tex}

\end{document}

%% file: lucivjansky_14.tex

\thispagestyle{plain}
\addcontentsline{toc}{subsection}{\numberline{}\hspace*{-15mm} 
N.~V.~Antonov, M.~Hnati\v{c}, A.~S.~Kapustin, T.~Lu\v{c}ivjansk\'y, L.~Mi\v{z}i\v{s}in:
\emph{Directed Percolation Process Advected by Compressible Velocity Field}}

\markboth{\sc T.~Lu\v{c}ivjansk\'y et al.} {\sc Directed Percolation Process}

\STM

\title{Directed Percolation Process Advected by Compressible Velocity Field}

\authors{N.~V.~Antonov$^{1}$, M.~Hnati\v{c}$^{2,3}$, A.~S.~Kapustin$^{1}$, 
T.~Lu\v{c}ivjansk\'y$^{3,4}$, L.~Mi\v{z}i\v{s}in$^{3}$
}

\address{$^{1}$
Department of Theoretical Physics, St. Petersburg 
University,\\Ulyanovskaya 1, St. Petersburg, Petrodvorets, 198504 Russia\\
$^{2}$
Institute of Experimental Physics, Slovak Academy of Sciences,\\
Watsonova 47, 040 01 Ko\v{s}ice, Slovak Republic\\
$^{3}$
Faculty of Sciences, P.J. \v{S}afarik University, \\
\v{S}rob\'arova 2, 041 54 Ko\v{s}ice, Slovakia\\
$^{4}$
Fakult\"at f\"ur Physik, Universit\"at Duisburg-Essen,\\
Lotharstrasse 1, 47048 Duisburg, Germany
}

\bigskip

\vspace*{3mm}

\begin{abstract}
The directed percolation process in the vicinity of non-equilibrium phase transition
is studied by the means of field theoretic methods.
It will be assumed that percolation takes place in a compressible environment, which
will be generated by the modified Kraichnan model.  We will discuss differences with the incompressible
case and describe how the given model can be renormalized. The renormalization constants
will be given to the one-loop order.
\end{abstract}
\section{Introduction \label{sec:intro}}
Nearly all physical phenomena are due
to non-equilibrium processes. Examples include chemical reactions in organisms, 
 heat transport in the atmosphere, social-economical systems and so on. 
 Despite the considerable effort the general principles
for far-from-equilibrium statistical mechanics are still missing.
Although general study of such systems is very demanding, there exist
situations, which are amenable to theoretical analysis.
 Analogously to the equilibrium critical behavior,
in the vicinity of some special points the scale invariance emerges that
can be handled within renormalization group (RG) technique.
The famous prototype is the directed percolation (DP) process \cite{StauAha},
which could be thought either as a kind of epidemic process\cite{Hin01} or as
Reggeon field theory in particle physics\cite{Cardy80}. 
The phase transition \cite{Hin01} in these systems 
occurs between the absorbing state (no infected individuals) and 
active state (nonzero number of infected individuals).
   
The deviations from the ideality are known to have a profound
impact. Immunization, long-range interactions or quenched disorder change the
critical behavior and in general give rise to new universality class \cite{Jan99,Hin07}. 

The aim of this paper is to study the directed percolation process in 
the presence of compressible advective environment.
Environment will be modeled by the modified Kraichnan model  \cite{Ant00, Ant06}, which allows
us to analyze effect of finite correlation time of velocity fluctuations
and of compressibility itself. 
\section{The model \label{sec:model}}
The continuum description of DP in terms of a density 
$\psi = \psi(t,\mx)$ of infected individuals typically arises from
a coarse-graining procedure in which a large number of
microscopic degrees of freedom were averaged out. Their
reminiscence is the presence of a Gaussian noise in the
corresponding Langevin equation.
The mathematical model must obviously respect the
absorbing state condition: $\psi = 0 $ is always a stationary state.
The coarse grained stochastic equation then reads \cite{JanTau04}
\begin{equation}
  \partial_t {\psi}  = D_0 (\boldnabla^2 - \tau_0)\psi  - 
   \frac{g_0 D_0}{2}\psi^2
  + \eta,
  \label{eq:basic}
\end{equation}
where $\eta$ denotes the noise term, $\partial_t = \partial / \partial t$ is
the time derivative, $\boldnabla^2$ is  the Laplace operator, $D_0$ 
is the diffusion constant, $g_0$ is the coupling constant and $\tau_0$ measures
 deviation from the threshold value for injected probability. The latter quantity
 can be thought
 as an analog to the temperature  variable in the standard $\varphi^4-$theory.
 \cite{JanTau04,Zinn}.
 Here and henceforth 
we distinguish unrenormalized (with subscript ``0'') quantities from renormalized ones
 (without subscript ``0''). The renomalized fields will be later denoted with the subscript $R$.

Although the rigorous proof is lacking, it is generally believed that
Langevin equation (\ref{eq:basic}) captures the gross properties
 of the percolation process and should contain relevant information about the
 large-scale behavior of the non-equilibrium phase
 transition between active $\psi > 0$ and absorbing state $\psi = 0$.   
Due to the absorbing state condition the correlator of $\eta $ has the following form
\begin{equation}
   \langle \eta(t_1,\mx_1) \eta(t_2,\mx_2) \rangle = g_0 D_0 \psi(t_1,\mx_1) 
   \delta(t_1-t_2) \delta^{(d)}(\mx_1 - \mx_2).
   \label{eq:noise_correl}
\end{equation}
The next step consists in the incorporation of the velocity fluctuations. The 
 standard way based on the
   the replacement $\partial_t$ by the Lagrangian derivative $\partial_t +({\bm v}\cdot\nabla)$ 
   is not sufficient. As was shown in \cite{AntKap10} the presence of compressibility 
   requires the following replacement
\begin{equation}
  \partial_t \rightarrow \partial_t +({\bm v}\cdot\nabla)+a_0 ({\bm \nabla}\cdot{\bm v}),
  \label{eq:subs}
\end{equation}
where $a_0$ is an additional parameter.
From the point of view of RG the introduction of $a_0$ is necessary,
 since it ensures multiplicative renormalizability of the model.
 
Our main aim here is to show how the influence of compressibility together with finite
 time correlated velocity field could be incorporated into the model of DP. To this end
we employ the model based on the modified
Kraichnan model \cite{Ant00,Ant99}. Such model properly describes main
features of the turbulent advection-diffusion propagation.

Following the work \cite{Ant00} we
 consider the velocity field to be random Gaussian variable with zero mean and 
 translational invariant correlator chosen in the form
\begin{equation}
  \langle v_i(t,{\bm x}) v_j (0,{\bm 0}) \rangle =
  \int \frac{{\mathrm d} \omega}{2\pi}
  \int \frac{{\mathrm d}^d {\bm k}}{(2\pi)^d} 
  D_v (\omega,\mk) {\mathrm e}^{-i\omega  t  +{\bm k}\cdot {\bm x}},
  \label{eq:vel_correl}
\end{equation}
where the kernel function $D_v(\omega,\mk)$ is given as
%
\begin{equation}
  D_v (\omega,\mk) = [P_{ij}^{k} + \alpha Q_{ij}^{k}]
  \frac{g_{10} u_{10} D_0^3 k^{4-d-y-\eta}}{\omega^2 + u_{10}^2 D_0^2 (k^{2-\eta})^2}.
  \label{eq:kernelD}
\end{equation}
Here $P_{ij}^k = \delta_{ij}-k_ik_j/k^2$ is transverse  and $Q_{ij}^k$ longitudinal
projection
operator, $k=|\mk|$, and $d$ is the 
dimensionality of the $\mx$ space.
Positive parameter $\alpha>0$ can be interpreted as a
deviation from the incompressibility condition ${\bm \nabla}\cdot {\bm v} = 0$.
The case $\alpha=0$ was studied previously \cite{AntKap10,AntKap08,DP13}.
The coupling constant $g_{10}$ and exponent $y$ describe the equal-time velocity
correlator or, equivalently, the energy spectrum \cite{Frisch,Vasiliev}.
On the other hand, the constant $u_{10}>0$ and the exponent $\eta$ describe
the characteristic frequency of the mode $k$.

The exponents $y$ and $\eta$ are analogous to the standard expansion parameter
$\varepsilon = 4-d$ in the static critical phenomena.
Later on it will be shown (Sec. \ref{sec:can_dim}) that the upper critical dimension of the pure
percolation problem is indeed $d_c = 4$. Therefore we retain the standard notation for the
exponent $\varepsilon$.
According to the general rules \cite{Zinn,Vasiliev} of RG approach we formally assume
that  the exponents $\varepsilon,y$ and $\eta$ are of the same order of magnitude and
in principle they constitute small expansion parameters in a perturbation sense.

%
%
%
The general form of kernel function in (\ref{eq:kernelD}) contains
various special limits. Their analysis is usually simpler and thus
 allows us to gain deeper physical insight.
 The possible limits are
\begin{enumerate}[i)]
  \item Rapid-change model, which is obtained in
        the limit $u_{10} \rightarrow \infty, g_{10}' \equiv g_{10}/u_{10} = const$.
	For this case we have
	$ D_v(\omega,{\bf k}) \propto g_{10}' D_0 k^{-d-y+\eta}$,
	and therefore the velocity field is now $\delta-$correlated in time variable.
  \item 'Frozen' velocity field: given by the limit $u_{10} \rightarrow 0$ and corresponds
	to the following form of Kernel function
	$ D_{v}(\omega,{\bf k}) \propto g_0
	D_0^2 \pi \delta(\omega) k^{2-d-y}$,
	In other words, the velocity field is quenched (time-independent).
  \item Pure potential velocity field:
	$\alpha\rightarrow\infty$ with $\alpha g_{10}=$constant. This limit is similar to
	the model of random walks in random environment with long-range
	correlations\cite{HonKar88}.
  \item Turbulent advection obtained setting $y = 8/3$ and $\eta=4/3$. This choise leads directly
        to the famous Kolmogorov ``five-thirds'' law for the spatial velocity correlations \cite{Frisch}.	
\end{enumerate}

For the effective use of RG method it is advantageous
to reformulate the stochastic problem (\ref{eq:basic}-\ref{eq:kernelD})
into the field-theoretic language. This can be achieved
 in the standard fashion
\cite{Vasiliev,deDom76,Janssen76}
and the resulting dynamic functional can be written as a sum of three terms
\begin{equation}
   \S[\varphi] = \S_{ \text{diff}}[\varphi]
   + \S_{\text{vel}}[\varphi]
   + \S_{\text{int}}[\varphi], 
   \label{eq:bare_act}
\end{equation}
where $\varphi=\{\tilde{\psi},\psi,\mv \}$ is the complete set of fields. In what follows
 $\psi^{\dagger}$ is the response field appearing after the noise field was integrated
 out \cite{Vasiliev}. 
The first term represents nothing else as free part of the equation (\ref{eq:basic})
and can be written as follows
\begin{equation}
  \S_{ \text{diff}}[\varphi] =  
    \tilde{\psi}[
  \partial_t - D_0\boldnabla^2+D_0\tau_0
  ]\psi ,
  \label{eq:act_diffuse}
\end{equation}
where the necessary integrations over time-spatial variables were omitted. For example
 second term stands for 
 \begin{equation}
   \tilde{\psi} \boldnabla^2\psi  =  
   \int \dRM t \int \dRM^{d} \mx \biggl\{
   \tilde{\psi}(t,\mx)\boldnabla^2\psi(t,\mx) \biggl\}.  
   \label{eq:example}
 \end{equation}
Since velocity fluctuations are considered as Gaussian stochastic variables, 
 the averaging procedure for them corresponds to the
functional integration with quadratic functional 
\begin{eqnarray}
  \S_{\text{vel}}[\mv] & = & -\frac{1}{2} 
  \mv_i D_{ij}^{-1} \mv_j,
  \label{eq:vel_action}
\end{eqnarray}
where $D_{ij}^{-1}$ is the kernel of the inverse linear operation in (\ref{eq:vel_correl})
and summation over repeated indices is implied.
The interaction part can be written as
\begin{eqnarray}
  \S_{\text{int}}[\varphi] & = &
    \tilde{\psi}\biggl\{  
  \frac{D_0\lambda_0}{2} [\psi-\tilde{\psi}
  ]
  -\frac{u_{20}}{2D_0}
  \mv^2 
  + (\mv\cdot\boldnabla) 
  +a_0  (\boldnabla\cdot\mv)
  \biggl\}\psi.
  \label{eq:inter_act}
\end{eqnarray}

All but the third term in (\ref{eq:inter_act}) stems directly from the nonlinear
parts in eqs. (\ref{eq:basic}) and (\ref{eq:subs}).
The third term proportional to $\propto \tilde{\psi}\psi\mv^2$ deserves a special consideration. 
Presence of such term is prohibited in the original Kraichnan model due
to the underlying Galilean invariance. However in our case the general form of
the velocity kernel function doesn't lead to such restriction. Also by direct
inspection of the perturbative RG method, one can show that such term will indeed be generated
(Consider first three Feynman graphs in the expansion (\ref{eq:exp_ppvv})). Such term was also
introduced in our previous work \cite{DP13}, where the incompressible analog
was studied. It was shown that such term doesn't lead to the significant differences
with respect to the quantitative values of universal quantities.

Basic ingredients of stochastic theory, correlation and response functions of the concentration 
field $\psi(t,\mx)$, can be in principle
computed as functional averages with respect to the weight functional $\exp(-\S)$.
Further the field-theoretic formulation summarized in (\ref{eq:act_diffuse})-(\ref{eq:inter_act})
 has the additional advantage, namely it is amenable to the full machinery of (quantum) field theory
 \cite{Zinn,Vasiliev}.
In subsequent section we will apply RG perturbative technique \cite{Vasiliev} that allows us
to study the model in the vicinity of its upper critical dimension $d_c=4$.
\section{Renormalization group analysis \label{sec:RG_analysis}}
The important goal of statistical theories is the determination of correlation and response functions
(usually called Green functions) of the dynamical fields as functions of the space-time coordinates. 
Graphically one can represent these functions in a form of sums over Feynman diagrams \cite{Zinn}.
Near criticality $\tau=0$ large fluctuations on all spatio-temporal scales dominate
the behavior of the system, which in turn results into the divergences in Feynman graphs.
The RG technique allows us to deal with them and at the same time
it serves us as an efficient technique for
determination of possible large scale behavior. Also it provides us
with the perturbative algorithm for estimation of universal quantities in the form
of formal expansion around upper critical dimension.

The critical behavior of the model is analyzed using standard RG procedure \cite{Vasiliev}. 
This procedure requires 
 action to be multiplicatively renormalizable and this goal can be achieved by 
 adding a new term $\psi^{\dagger}\psi vv$ into the total action with new 
 independent parameter (charge) $u_2$.
\subsection{Canonical dimensions \label{sec:can_dim}}
The analysis of ultraviolet (UV) divergences is based on the power 
counting procedure. This step allows us to identify UV divergent
 structures in the perturbation theory.
For translational invariant
systems it  is sufficient to analyze only 1-particle irreducible (1PI)
graphs \cite{Zinn,Vasiliev}.

In contrast to the static models, dynamic models contain two independent scales: frequency scale 
and momentum scale.
The corresponding dimensions for each quantity can be found using the 
standard normalization conditions 
\begin{eqnarray}
  & & d_k^k = - d^k_x =1,\quad
  d^k_\omega=d_t^k = 0,\nonumber\\
  & & d_k^\omega = d^\omega_x = 0,\quad
  d^\omega_\omega = -d_t^\omega = 1
  \label{eq:def_normal}
\end{eqnarray}
together with condition to field-theoretic action (\ref{eq:bare_act})
be a dimensionless quantity.
Further based on the values $d^\omega_Q$ and $d_Q^k$, one can introduce
the total canonical dimension $d_Q$
\begin{equation}
   d_Q = d_Q^k + 2d_Q^\omega,
\end{equation}
whose form can be obtained from the comparison of IR most
relevant terms ($\partial_t \propto \boldnabla^2$) in the action.

The dimensions of all quantities for the model are summarized in Table \ref{tab:canon}.
It follows that the model is logarithmic (when coupling constants
are dimensionless) at $\varepsilon = y = \eta = 0$, and the UV divergences
have the form of the poles in these small parameters.
\begin{table}
\centering
\begin{tabular}{| c | c | c | c | c| c | c | c | c | }
  \hline
  $Q$ & $\psi,\tilde{\psi}$ & ${\mv}$ & $D_0$ & $\tau_0$ & $g_{10}$ & $\lambda_0 $  
  & $u_{10}$  & $u_{20},a_0,\alpha$
  \\ \hline
  $d_Q^k$ & $d/2$ & $-1$ & $-2$ & $2$ & $y$ & $\varepsilon/2$
  & $\eta$  & $0$ 
  \\ \hline 
  $d^\omega_Q$ & 0 & $1$ & $1$ & $0$ & $0$ & $0$
  & $0$  & $0$ 
  \\ \hline 
  $d_Q$ & $d/2$ & $1$ & $0$ & $2$ & $y$ & $\varepsilon/2$
  & $\eta$  & $0$ 
  \\ \hline    
\end{tabular}
  \caption{Canonical dimensions of the fields and parameters 
	  for the model (\ref{eq:act_diffuse})-(\ref{eq:inter_act}).
  }
  \label{tab:canon}
\end{table}
The total dimension $d_Q$ plays for the dynamical models
the same role as does the conventional (momentum) dimension in static problems.
The total canonical dimension of an arbitrary 1PI Green function
is given by the relation
\begin{equation}
   d_\Gamma = d^k_\Gamma + 2 d^\omega_\Gamma = d + 2 - \sum_i N_i d_i, \quad 
   i\in\{\tilde{\psi}, \psi, \mv \}
\end{equation}
Using this relation one immediately obtains the divergent functions listed in
Tab. \ref{tab:functions}.
\begin{table}
\centering
\begin{tabular}{| c | c | c | c | c| c | }
  \hline
  $\Gamma_{\text{1PI}}$ & $\Gamma_{\psi,\tilde{\psi}}$ & 
  $\Gamma_{\tilde{\psi}\psi\mv}$ & $\Gamma_{\tilde{\psi}^2\psi}$ &$\Gamma_{\tilde{\psi}\psi^2}$ &
  $\Gamma_{\tilde{\psi}\psi \mv^2}$   
  \\ \hline
  $d_\Gamma$ & $2$ & $1$ & $\varepsilon$ & $\varepsilon$ & $0$
  \\ \hline   
\end{tabular}
  \caption{Canonical dimensions of the potentially divergent 1PI functions.
  }
  \label{tab:functions}
\end{table}
\subsection{Computation of the RG constants \label{sec:RG_const}}
Here, we will summariz the main steps of the perturbative aspects of the RG approach.
Perturbation theory is developed starting from the free part of the action, which are graphically
represented as lines in the Feynman diagrams. On the other hand the non-linear terms  in
(\ref{eq:inter_act}) play the role of vertices.

For the calculation of RG constants we have used dimensional regularization in the
minimal subtraction scheme (MS). In this scheme the expansions
parameters are  $\varepsilon,y$ und $\eta$ and the poles in renormalization constants
can be realized in the form of their linear combination.
Because the finite correlated case involves two different dispersion laws: 
$\omega\propto k^2$ for the scalar and
$\omega\propto k^{2-\eta}$
for the velocity fields, the calculations for the renormalization constants become
rather cumbersome already in the one-loop approximation \cite{Ant00,Ant99}.
However, as was shown in \cite{AdzAntHon02} to the two-loop order it is sufficient to
consider $\eta = 0$. This choice substantially simplifies the practical calculations
and as can be seen from the explicit expressions (\ref{eq:RG_inverse1}), the
only poles to the one-loop order are $1/\varepsilon$ and $1/y$. We stress that this simple picture
pertains only to the lowest orders in
perturbation scheme. In higher order terms poles in the form
of general linear combinations in $\varepsilon,\eta$ and $y$ are expected to arise.

\begin{figure}
\begin{tabular}{ c c c }
    \includegraphics[width=2cm]{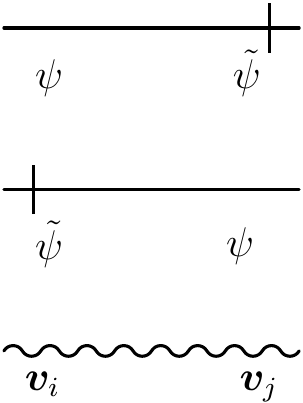}      
 &
    \includegraphics[width=4.5cm]{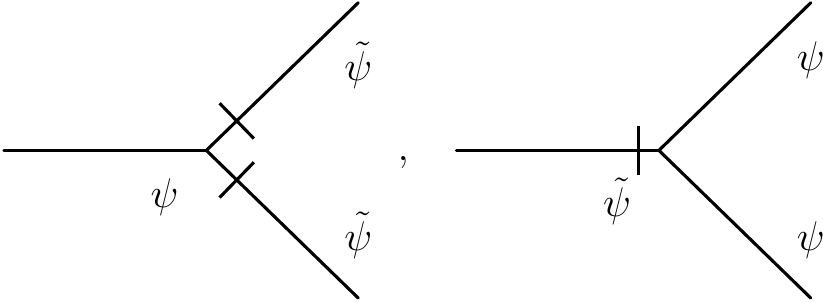}    
 & 
    \includegraphics[width=4.5cm]{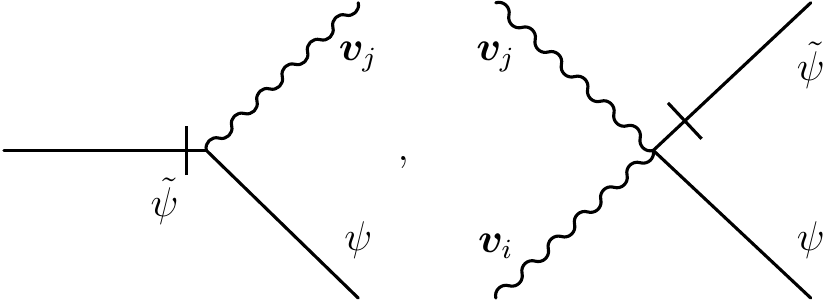}    
  \end{tabular}
\caption{Diagrammatic representation of the elements of perturbation theory.}
 \label{fig:prop}
 \end{figure}

The perturbation theory of the model (\ref{eq:bare_act}) is amenable
to the standard Feynman diagrammatic expansion \cite{Zinn,Vasiliev}.
  Studied model  contains three different types of propagators
and they are easily read off from the Gaussian part of the model
given by (\ref{eq:act_diffuse}) and (\ref{eq:vel_action}), respectively. 
Their graphical representation
is depicted in Fig.~\ref{fig:prop}. The corresponding algebraic expressions 
 are
\begin{eqnarray}
   \langle {\psi} \tilde{\psi}\rangle_0 
   & = &
    \langle \tilde{\psi} \psi\rangle_0^* 
    =
   \frac{1}{-i\omega + D(k^2+\tau)},\\
   \label{eq:prop1a}
   \langle \mv \mv \rangle_0 
   & = & 
   [P_{ij}^{k} + \alpha Q_{ij}^{k}]
   \frac{g_{10} u_{10} D_0^3 k^{4-d-y-\eta}}{\omega^2 + u_{10}^2 D_0^2 (k^{2-\eta})^2}.
   \label{eq:prop2a}
\end{eqnarray}

The interaction vertices from the nonlinear part of the action
(\ref{eq:inter_act}) describe
 the fluctuation effects connected with the 
 percolation processed itself,
 advection of concentration field and the interactions between
 the velocity components, respectively.
With every such vertex
the so-called vertex factor \cite{Vasiliev} 
\begin{equation*}
  V_N(x_1,\ldots,x_N;\varphi) = 
  \frac{\delta^N \S_{\text{int}}[\varphi]}{\delta\varphi(x_1)\ldots\delta\varphi(x_N)},
  \quad
  \varphi \in\{\tilde{\psi},\psi,\mv \}
  \label{eq:ver_factor}
\end{equation*}
is associated. 
In our model we have to deal with four interaction vertices, which are
also graphically depicted in the Fig.~\ref{fig:prop}.
The corresponding vertex factors are
\begin{equation}
    V_{\tilde{\psi}{\psi} \psi} =
       - V_{\tilde{\psi}\tilde{\psi} \psi} = D_0\lambda_0,\quad
   V_{\tilde{\psi}\psi\mv} = - \frac{u_{20}}{D_0},\quad
    V_{\tilde{\psi}\psi\mv\mv} = ik_i +i a_0 q_i,
   \label{eq:factors}
\end{equation}
where in the last terms $k_i$ is the momentum of the field $\psi$ and $q_i$
is the momentum carried by the velocity field $\mv$.
The presence of the interaction vertex $V_{\tilde{\psi}\psi\mv\mv}$ leads
to the proliferation of the new Feynman graphs, which
were absent in the previous studies \cite{AntKap10,AntKap08,DP13}. 
%
By direct inspection of the Feynman diagrams one can easily 
observe that the real expansion parameter is 
rather $\lambda_0^2$ than $\lambda_0$. This is a consequence of the duality symmetry 
of the action for the pure percolation problem with respect to time reversal
$ \psi(t,\mx) \rightarrow -\tilde{\psi}(-t,\mx),\quad
  \tilde{\psi}(t,\mx) \rightarrow -\psi(-t,\mx)$.
Therefore it is advantageous to introduce
new charge $g_{20}$ by as $g_{20} = \lambda_0^2$.  
We note that in the presence of compressible velocity field this transformation
has to be supplemented by additional transformation $ a_0 \rightarrow 1-a_0$
  as can be directly seen.
Using Tab. \ref{tab:canon} we introduce the renormalized parameters via
\begin{align}
   \label{eq:RGconst}
   &D_0 = D Z_D, &\tau_0& = \tau Z_\tau + \tau_c,
     &a_0& = a Z_a,  
     \nonumber \\ 
   &g_{10} = g_{1} \mu^{y+\eta} Z_{g_1}, &u_{10}& = u_1 \mu^\eta Z_{u_1},
   &\lambda_0& = \lambda \mu^{\varepsilon} Z_\lambda,
   \\
   &g_{20}=g_2 \mu^{2\varepsilon} Z_{g_2}, &u_{20}& = u_2 Z_{u_2}, \nonumber
\end{align}
where $\mu$ is the reference mass scale in the MS scheme~\cite{Zinn} and $\tau_c$
stands for fluctuation shift of the critical point.
The choice (\ref{eq:RGconst}) together with 
renormalization of fields $ \tilde{\psi} = Z_{\tilde{\psi}} \tilde\psi_{R},\quad
 \psi = Z_\psi \psi_{R},\quad \mv = Z_v\mv_{R} $
is sufficient to have fully UV renormalized theory.
The total renormalized action for
the renormalized fields $\varphi_R \equiv \{\tilde\psi_{R}, \psi_{R}, \mv_{R} \}$ 
  can be written in the compact form
\begin{eqnarray}
  \S_R[\varphi_R] & = & \
   \tilde{\psi}_R \biggl[
  Z_1\partial_t - Z_2 D\nabla^2 + Z_3 D\tau 
   +  Z_4 (\bm{v}_R\cdot \bm{\nabla}) 
    +  a Z_5 (\bm{\nabla}\cdot\bm{v}_R) 
   \biggl] \psi_R \nonumber\\
   &  - &
   \frac{D\lambda}{2}[Z_6 \tilde{\psi}_R 
   -  Z_7
  \psi_R]\tilde{\psi_R}\psi_R - Z_8\frac{u_2}{2D}\tilde{\psi_R} \psi_R \bm{v_R}^2 +
   \frac{\bm{v}_R D_{Rv}^{-1}\bm{v}_R}{2}.
   \label{eq:renorm_action}
\end{eqnarray}
The relations between 
renormalization constants can be directly read off from the action (\ref{eq:renorm_action}), which yields
\begin{align}
  \label{eq:RG_direct}
  &Z_1  = Z_\psi Z_{\psi^\dagger}, 
  & Z_2 & =  Z_\psi Z_{\psi^\dagger} Z_D, \nonumber \\
  &Z_3  = Z_\psi Z_{\psi^\dagger} Z_D Z_\tau,
  & Z_4 & =  Z_\psi Z_{\psi^\dagger} Z_v, \nonumber \\
  &Z_5  = Z_\psi Z_{\psi^\dagger} Z_v Z_a, 
  & Z_6 & =  Z_\psi Z_{\psi^\dagger}^2 Z_D Z_\tau, \\
  &Z_7  = Z_\psi^2 Z_{\psi^\dagger} Z_D Z_\lambda, 
  & Z_8 & =  Z_\psi Z_{\psi^\dagger} Z_v^2 Z_{u_2}Z_D^{-1}. \nonumber  
\end{align}
After the theory is made UV finite through the appropriate choice
of RG constants $Z_1,\ldots,Z_8$ one can use the relations (\ref{eq:RG_direct})
to obtain corresponding RG constants for the fields and parameters
appearing in (\ref{eq:RGconst}). 
Note that according to the general rules of RG the nonlocal terms should not be 
renormalized. From the inspection of kernel function (\ref{eq:kernelD})
one thus obtain two additional relations $  1  = Z_{u_1} Z_D$ and
  $1  =  Z_{u_1} Z_{g_1}Z_D^3 Z_v^{-2}$.  
  
Now we give a brief overview of how the renormalization constants
were computed. The number
of divergent Feynman graphs do not pose a serious problem to the first order of the
perturbation theory. Moreover their analysis is to some extent simplified by the two facts:
\begin{enumerate}
 \item Integral of a power of internal momenta is zero in dimensional regularization. 
       Hence the tadpole diagrams are immediately discarded.
 \item Closed circuits of propagators $\tilde{\psi}\psi$ vanish identically, which 
       is a consequence of the {It\^o} time discretization\cite{Vasiliev}
       that we have here considered.	  
\end{enumerate}
From the two-point Green functions only $\Gamma_{\tilde{\psi}\psi}$ deserves
special attention. For it one can write down the following
Dyson equation
\begin{eqnarray}
  \Gamma_{\tilde{\psi}\psi}
  & =& i\omega Z_1 - D p_2 Z_2
  -D\tau Z_3 + 
  \raisebox{-0.25ex}{ \includegraphics[width=2cm]{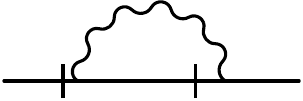}}
    + 
   \frac{1}{2} \raisebox{-0.25ex}{ \includegraphics[width=2cm]{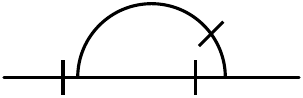}   }.
  \label{eq:exp_pp}
\end{eqnarray}
The perturbation expansion for the interaction vertices can be successively presented
in the following way:
 \begin{eqnarray}
  \langle \tilde{\psi} {\psi} \mv \rangle_{1PI} 
   \Gamma_{\tilde{\psi} {\psi} \mv}
   & = & 
   -ip_j Z_4 - ia q_j Z_5 +   
   \raisebox{-5.25ex}{\includegraphics[width=2cm]{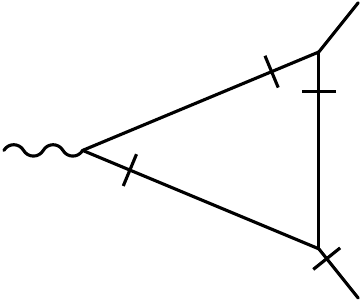}} 
      +  \frac{1}{2}
  \raisebox{-5.25ex}{ \includegraphics[width=2cm]{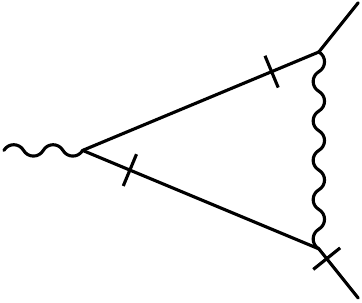}} \nonumber\\
    & + & \raisebox{-3.5ex}{ \includegraphics[width=2cm]{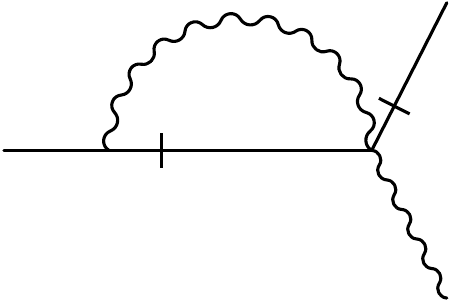}}
     +  \raisebox{-3.5ex}{ \includegraphics[width=2cm]{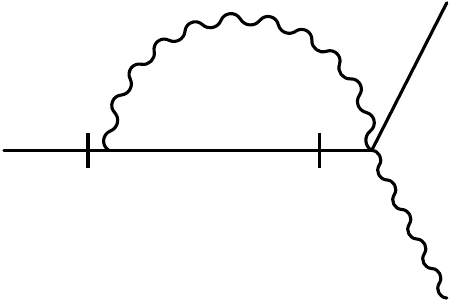}},
   \label{eq:exp_ppv}
 \end{eqnarray}
 \begin{eqnarray}
   \Gamma_{\tilde{\psi}\tilde{\psi} \psi}
   & =& D\lambda Z_6 +
   \raisebox{-5.25ex}{ \includegraphics[width=2cm]{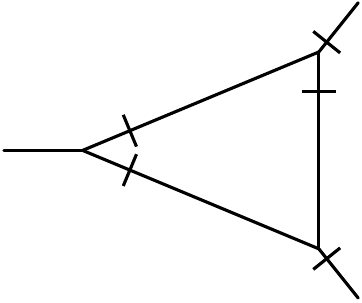}} 
    + 
   \frac{1}{2}
   \raisebox{-5.25ex}{ \includegraphics[width=2cm]{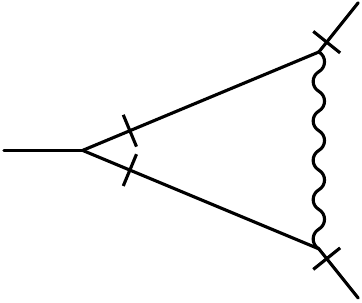}} 
     + 
   \raisebox{-5.25ex}{ \includegraphics[width=2cm]{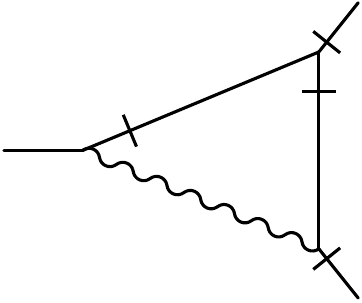}},  
   \label{eq:exp_ppp1}
 \end{eqnarray}
 \begin{eqnarray}
   \Gamma_{\tilde{\psi}{\psi} \psi}
   & = & -D\lambda Z_7 +
   \raisebox{-5.25ex}{ \includegraphics[width=2cm]{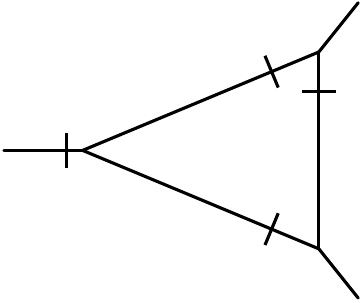}} 
    + 
   \frac{1}{2}
   \raisebox{-5.25ex}{ \includegraphics[width=2cm]{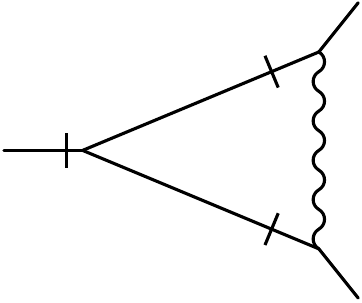}} 
     + 
   \raisebox{-5.25ex}{ \includegraphics[width=2cm]{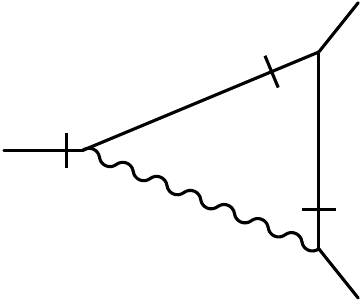}},
   \label{eq:exp_ppp2}
 \end{eqnarray}
 \begin{eqnarray}
   \Gamma_{\tilde{\psi}{\psi} \mv \mv}
   & = & \frac{u_{2}}{D}\delta_{ij} Z_8 +
   \raisebox{-3.75ex}{ \includegraphics[width=1.5cm]{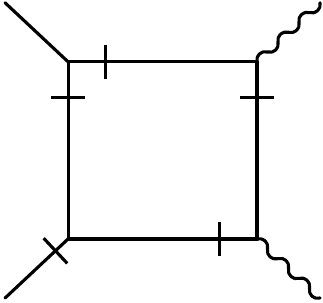}} 
    +   
   \raisebox{-3.75ex}{ \includegraphics[width=1.5cm]{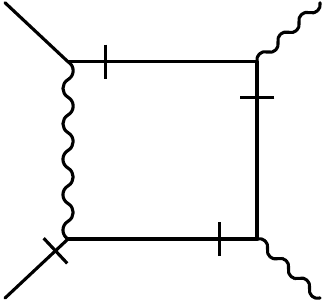}} 
     +  \frac{1}{2}
   \raisebox{-3.75ex}{ \includegraphics[width=1.5cm]{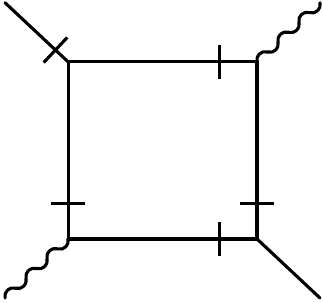}}  
   + \raisebox{-4.5ex}{ \includegraphics[width=2cm]{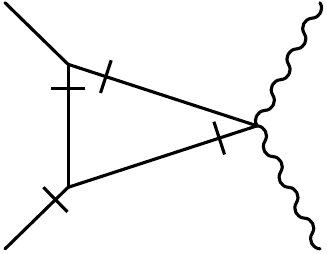}}  
   \nonumber \\
   & + &
   \raisebox{-4.5ex}{ \includegraphics[width=2cm]{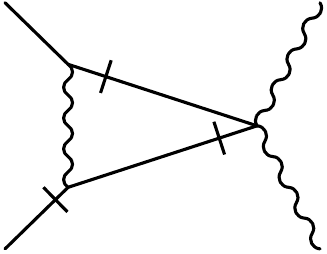}}  
   + \raisebox{-4.5ex}{ \includegraphics[width=2cm]{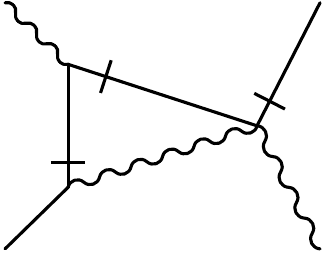}}  
    + 
   \raisebox{-4.5ex}{ \includegraphics[width=2cm]{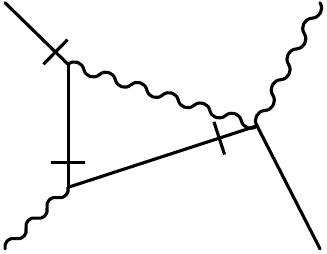}}  
   + \raisebox{-4.5ex}{ \includegraphics[width=2cm]{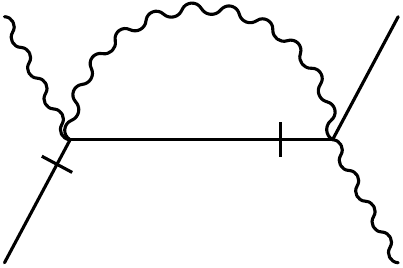}}.
   \label{eq:exp_ppvv}
\end{eqnarray}
For  the graphs whose symmetry coefficient \cite{Vasiliev} is different from $1$, we have
explicitly stated its numerical value.
Note again that in the language of Feynman graphs 
the need for the term $\propto \tilde{\psi}\psi\mv^2$ can be traced out
to the first three Feynman graphs in (\ref{eq:exp_ppvv}), which in the
sum doesn't cancel out as is the case for incompressible velocity field.

The computation of the diverging parts of the Feynman graphs
follows the usual methods
of dimensional regularization \cite{Vasiliev} and the 1-loop results
are
\begin{align}
   \label{eq:RG_inverse1}
   Z_1 & =  1 + \frac{g_1\alpha a(1-a)}{(1+u_1)^2y} + \frac{g_2}{4\epsilon},
   \nonumber  \\   
   Z_2 & =  1 - \frac{g_1}{4(1+u_1)y}\biggl[ 3+ 
     \alpha \NB{\frac{u_1-1}{u_1+1}-\frac{4a(1-a)u_1}{(1+u_1)^2} } \biggl]
    +  \frac{g_2}{8 \epsilon},
    \nonumber \\
   Z_3 & =  1 + \frac{g_1\alpha a(1-a) }{(1+u_1)^2 y} +\frac{g_2}{2 \epsilon},
   \nonumber \\
   Z_4 & =  1 + \frac{g_1}{4(1+u_1)^2 y}\biggl[\alpha
   \NB{1+\frac{4a(1-a)u_1}{1+u_1} } 
   -  u_2(6+6u_1+2\alpha u_1)  \biggl]  + \frac{g_2}{4\epsilon},
    \nonumber
   \\
   Z_5 & =  1 + \frac{g_1\alpha}{4(1+u_1)^2y}\biggl[1+2(1-a)
   \NB{2a-\frac{1}{1+u_1} }
   \biggl]
   \nonumber\\
   & - \frac{g_1 u_2}{4a(1+u_1)y}\biggl[ 3+\alpha - \frac{2\alpha(1-a)}{1+u_1}
   \biggl] + \frac{g_2(4a-1)}{8a\epsilon},
   \nonumber \\
   Z_6 & =  1 - \frac{g_1\alpha(1-a) }{(1+u_1) y}\biggl[1-a-\frac{2a}{1+u_1} \biggl]
    + \frac{g_2}{\epsilon},
    \nonumber \\
   Z_7 & =  1 - \frac{g_1\alpha a}{(1+u_1) y}\biggl[
   a-\frac{2(1-a)}{1+u_1}
   \biggl] + \frac{g_2}{\epsilon},
   \nonumber \\
   Z_8 & =  1 +\frac{g_1}{2(1+u_1)y}\biggl[\alpha\frac{2a(1-a)+1}{1+u_1} -
   \frac{\alpha a(1-a)}{u_2(1+u_1)^2} -  u_2(3+\alpha) \biggl] + \frac{g_2}{2\epsilon}.
\end{align}
The ubiquitous geometric factors stemming from the angular integration
were included into the renormalized charges $g_1$ and $g_2$ via the
redefinitions:$g_1/16\pi^2 \rightarrow g_1, g_2/16\pi^2 \rightarrow g_2$.
The equations (\ref{eq:RG_inverse1}) must satisfy certain conditions
dictated by the aforementioned time reversal symmetry. This symmetry results into the
following conditions \cite{AntKap10} for the renormalization constants
\begin{align}
   \label{eq:RG_cond}
   & Z_i(a) = Z_i(1-a)\quad i\in\{1,2,3,4,8 \}, \nonumber \\
   & Z_6(a) = Z_7(1-a), \quad Z_7(a) = Z_6(1-a), \nonumber \\
   & Z_1(a) - aZ_5(a) = (1-a) Z_5(1-a), 
\end{align}
where the RG constants are considered as functions of renormalized 
parameter $a$.
These relations can be directly checked for our results (\ref{eq:RG_inverse1}).
Further the relations (\ref{eq:RG_direct}) could be inverted with respect
to the RG constants for the fields and parameters in a straightforward manner to yield
\begin{align}
  \label{eq:RG_inverse2}
  &Z_D = Z_2 Z_1^{-1},
  &Z_\tau& = Z_3 Z_2^{-1},
  &Z_v& = Z_4 Z_1^{-1}, \nonumber\\
  &Z_a = Z_5 Z_4^{-1},
  &Z_{\psi}& = Z_1^{1/2} Z_6^{-1/2} Z_7^{1/2},
  &Z_{\psi^\dagger}& = Z_1^{1/2} Z_6^{1/2} Z_7^{-1/2},\nonumber \\ 
  &Z_{u_1} = Z_1 Z_2^{-1},     
  &Z_{\lambda}& = Z_{1}^{-1/2} Z_2^{-1} Z_6^{1/2} Z_7^{1/2},
  & Z_{g_2}& = Z_1^{-1} Z_2^{-2} Z_6 Z_7,\nonumber \\
  & Z_{u_2}  = Z_2 Z_8 Z_4^{-2},
  &Z_{g_1}& = Z_2^{-2} Z_4^2. 
\end{align}
After insertion explicit results for renormalization constants
(\ref{eq:RG_inverse1}) one obtains desired RG constants.
\section{Fixed points and scaling regimes \label{sec:regimes}}
%
%
%
The scaling behavior in the infrared (IR) limit can be studied by analyzing RG 
flow as $\mu\rightarrow 0$ after the renormalization procedure to given order of perturbation
scheme is performed.
The possible IR asymptotic behavior is 
governed by the fixed points (FPs) of the $\beta$-functions \cite{Vasiliev}. 
The fixed points $g^{*}=\{g_1^{*},g_2^{*},u_1^{*},u_2^{*},a^{*}\}$ 
can be found from requirement that all $\beta$ functions  simultaneously vanish
\begin{equation}
  \beta_{g_1} (g^{*}) =\beta_{g_2} (g^{*})= \beta_{u_1}
  (g^{*})=\beta_{u_2} (g^{*})=\beta_{a} (g^{*})=0,
  \label{eq:gen_beta}
\end{equation}
where the $\beta$ functions, which express the flows of parameters under the RG 
transformation \cite{Zinn,Vasiliev}, are defined as $ \beta_g = \mu \partial_{\mu} g |_{0}$.
Here $\dots|_0$ denotes fixed bare parameters.
Whether the given FP could be realized in physical systems (IR stable) or not 
(IR unstable)
is determined by the eigenvalues of the matrix $\Omega=\{\Omega_{ij}\}$ with
the components $\Omega_{ij} = {\partial \beta_i}/{\partial g_j}$,
where $\beta_i$ stands for the full set of $\beta$ functions and $g_j$ is 
the full set of charges $\{ g_1 ,g_2 ,u_1 ,u_2 ,a \}$.
For the IR stable FP the real part eigenvalues of matrix $\Omega$ are
strictly positive. In general these conditions determine the region of stability for
the given FP in terms of $\varepsilon,\eta$ and $y$.    
Further one exploit fact that the bare Green functions are independent of $\mu$
to obtain RG equation.
Applying the differential operator $\mu\partial_\mu$ at fixed bare quantities leads to
 the following equation for the renormalized Green function $G_R$ 
\begin{equation}
  \{ D_{\text{RG}} + N_{\psi} \gamma_{\psi} + N_{\psi^{\dagger}} \gamma_{\psi^{\dagger}}
  + N_{\mv} \gamma_v
  \} 
  G_R(e, \mu, \dots)=0,
  \label{eq:basic_RG}
\end{equation}
where $e$ is the full set of renormalized counterparts of the bare
 parameters $e_0 =\{D_0, \tau_0, u_{10},u_{20}, g_{10}, g_{20}, a_0 \}$ and $\dots$  denotes 
 other parameters, such as spatial or time variables. 
 The RG operator $D_{\text{RG}}$ is given in the form
\begin{equation}
   D_{\text{RG}}\equiv \mu\partial_\mu|_0  = \mu \partial_{\mu} + 
            \sum_{g} \beta_g \partial_g 
	    - \gamma_D D_D - \gamma_{\tau} D_{\tau},
   \label{eq:RG_equation}	    
\end{equation}
where $g\in\{u_1,u_2,g_1,g_2,a \}$, $D_x = x \partial_x$ for any variable $x$, 
and $\gamma_x$ are anomalous dimensions of the quantity $x$ and
are defined as $ \gamma_x \equiv \mu\partial_\mu \ln  Z_x |_0$.
Application of 
this definition 
on the relations 
 (\ref{eq:RGconst}) leads in straightforward manner to the equations
\begin{align}
  &\beta_{g_1} = g_1 (-y + 2\gamma_D-2\gamma_v), 
    & \beta_{g_2}& = g_2 (-\epsilon -\gamma_{g_2}), 
     &  \beta_{u_1}& = u_1(-\eta +\gamma_D), \nonumber \\
  &\beta_{u_2} = - u_2 \gamma_{u_2}, &\beta_{a}& = - a \gamma_{a}.
  \label{eq:beta_functions}
\end{align}
Last equation suggests that either $a=0$ or $a\neq 0$ is satisfied. However, as the
explicit results (\ref{eq:RG_inverse1}) show parameter $a$ could also
appear in the denominator. Similar reasoning also applies for the function $\beta_{u_2}$.

From the structure of anomalous dimensions it is clear that the
resulting systems of equations for FPs is quite involved.
Hence in order to gain some physical insight into their
structure it would be reasonable to divide to overall analysis into some specials cases.

\section{Conclusions\label{sec:concl}}
 In this article we have summarized main points of the field-theoretic study
  of directed percolation process in the presence of compressible velocity field. We have obtained
  field theoretical action and presented renormalization procedure of the model to the one-loop order.
  The next step consists in detailed analysis of the possible scaling
  regimes, which however due to the complicated form of RG functions will be published elsewhere \cite{new}.
\subsection*{Acknowledgement}
The work was supported by VEGA grant No. $1/0222/13$ and No. $1/0234/12$
 of the Ministry of Education, Science, Research and Sport of the Slovak Republic. This article was also
created by implementation of the Cooperative phenomena and phase transitions
in nanosystems with perspective utilization in nano- and biotechnology project No
26110230097. Funding for the operational research and development program was
provided by the European Regional Development Fund.